\documentclass{emulateapj}

\shorttitle{E+A Galaxies with Blue Cores}
\shortauthors{Yang et al.}

\begin{document}

\title{E+A Galaxies with Blue Cores: Active Galaxies in Transition}

\author{Yujin Yang\altaffilmark{1}, Christy A. Tremonti\altaffilmark{1,2}, Ann I.\ Zabludoff\altaffilmark{1}, and Dennis Zaritsky\altaffilmark{1}}

\altaffiltext{1}{Steward Observatory, University of Arizona, Tucson, AZ 85721;  
yyang, tremonti, azabludoff, and dzaritsky@as.arizona.edu.}
\altaffiltext{2}{Hubble Fellow}

\begin{abstract}
\emph{HST} ACS images reveal blue cores in four E+A, or post-starburst, galaxies.
Follow-up spectroscopy shows that these cores have LINER spectra.
The existence of LINERs, consistent with those in many elliptical galaxies, is yet 
one more piece of evidence that these post-merger, post-starburst, bulge-dominated 
galaxies will evolve into normal ellipticals.  
More interestingly, if LINERs are powered by low-luminosity AGN, their presence 
in these E+As suggests that any rapid growth phase of the central black hole ended 
in rough concert with the cessation of star formation.
This result emphasizes the importance of E+As for exploring how the evolution of 
black holes and AGN may be tied to that of galactic bulges.
\end{abstract}


\section{Introduction}

The strong correlation between black hole mass and galaxy 
bulge velocity dispersion \cite[$M_\bullet-\sigma_B$;][]{Ferrarese00,Gebhardt00}
suggests one of two things.
Either early-type galaxies do not arise from dissipative mergers or 
there is a connection --- perhaps causal ---
between the small-scale physics of black hole (BH) growth and 
the large-scale physics that organizes the host galaxy morphology, 
kinematics, and stellar populations during and after the merger.
Given the observational evidence that gas-rich mergers do occur and that at least 
some produce pressure-supported, bulge-dominated remnants, 
it remains to understand how the processes 
that drive the evolution of the smallest and largest galactic scales are related.

A key to resolving this question is to identify galaxies undergoing 
large-scale transitions via mergers and to consider the properties of their cores.
On-going mergers are too complicated to provide clear answers, while their likely 
remnants are too far removed from the merger event.  
E+A galaxies, which have post-starburst spectra, frequent tidal features, 
and the kinematic and morphological signatures of
early type galaxies \citep{Zabludoff96,Norton01,Chang01,Yang04},
are true transitional objects and thus plausible test cases. 
In this {\it Letter}, we explore whether there is evidence for a central BH/AGN 
in nearby E+As, and, if so, whether the core is
consistent with those of early-type galaxies and 
evolving in concert with the galaxy as a whole.

\section{Data: Imaging and Spectroscopy}

\begin{figure}
\epsscale{0.90}
\plotone{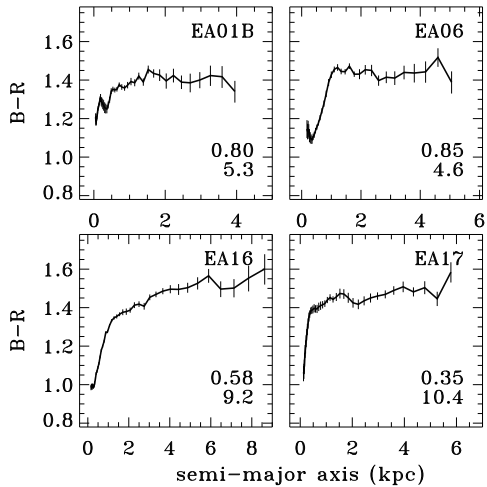}
\caption{
	($B-R$) color profiles of four E+A galaxies with blue cores.
	We	note the bulge fractions (bulge-to-total light) and S\'ersic indices $n$
	(from a single S\'ersic $r^{1/n}$ profile fit)
   derived from the GALFIT algorithm \citep{Peng02}.
	The spheroid-dominated morphologies and post-starburst spectra are 
	consistent with these galaxies being 
	in transition from late- to early-types.} 
\label{fig:image}
\end{figure}

Our sample
is a subset of the 20 nearby ($0.06 < z < 0.12$) E+As 
(defined to have strong Balmer absorption lines,
$\langle \rm{H} \beta \gamma \delta \rangle > 5.5 \  \AA$,
but little or no [\ion{O}{2}] emission, EW[\ion{O}{2}] $<$ 2.5 \AA)
in the Las Campanas Redshift Survey \citep{Zabludoff96}.
Initial inspection of our \emph{HST}
WFPC2 and ACS  $B$ and $R$ images\footnotemark\ 
reveals at least six galaxies with bright, blue, almost stellar-like cores
\citep{Yang06}.
In Fig. \ref{fig:image}, we show 
the color profiles for three of these galaxies (EA06, EA16, EA17) and 
for another blue-core galaxy (EA01B) that was observed serendipitously.
EA01B is the disturbed companion galaxy of EA01A \cite[originally EA1 in][]{Zabludoff96}
that we have now determined spectroscopically to lie at the same redshift 
as EA01A and also to have an E+A spectrum. The EA01A-B system is the first
known binary E+A system and provides additional evidence 
that the E+A phase of galaxy evolution is triggered by galaxy-galaxy interactions.
The central blue core of each galaxy has $\Delta(B-R)\simeq0.3$ with respect 
to the outer galaxy and a characteristic size of 1 kpc.
\footnotetext{($B, R$) corresponds to (F435W, F625W) for the ACS imaging (EA06, EA16, 
and EA17) 	and to (F439W, F702W) for the WFPC2 imaging (EA01), respectively.}
With the exception of the blue cores, the morphologies of these galaxies
are consistent with elliptical/S0 galaxies (EA01B, EA06, EA16) 
or early spiral galaxies (EA17).

The blue cores, while too faint to extract independently from
models of the underlying galaxy light, raise the possibility of a central AGN
\citep[e.g.,][]{O'Connell05} or of a luminous, compact star cluster 
\citep[e.g.,][]{Colina_et_al_2002}.
We therefore
obtained long-slit spectra of the four blue-core E+As (and EA01A)
with the Inamori Magellan Areal Camera and Spectrograph (IMACS) 
on the Magellan Baade 6.5 meter telescope between April 3 and 5, 2005.
We used the 300 line mm$^{-1}$ grism and $0.7\arcsec$ wide slit, which resulted 
in a dispersion of 1.34 \AA\ pixel$^{-1}$ and a spectral resolution of $\sim$ 4.2 \AA\
over the wavelength range 4000-8000 \AA.  We took three 600s or 900s exposures 
for each galaxy to remove the cosmic rays.
The frames were processed and flux-calibrated in the standard manner with IRAF.
The following analysis uses one-dimensional
spectra extracted from a $0.7\arcsec \times 2.0\arcsec$ aperture.
No X-ray data are yet available for these sources to help
discriminate between the possible origins of the blue cores.

\section{Nebular Diagnostics of Nuclear Activity}
\label{sec:fitting}
Optical nebular emission lines provide key diagnostics of the
state of the interstellar medium and the source of ionization (AGN or hot stars).
Measuring these lines presents a particular challenge in E+As because
they are weak (by definition) and superposed on a complex stellar
continuum.  Conventional continuum subtraction techniques are
inadequate, especially for the Balmer emission lines, which can be
swallowed by 5~\AA\ or more of stellar absorption.  Instead, we fit a
stellar population model to the continuum following
\citet{Tremonti_et_al_2006}. We use a library of template spectra
drawn from the stellar population models of
\citet{Bruzual_and_Charlot_2003}, which have a spectral resolution of
$\sim3$~\AA.  We use templates spanning a range of ages (1, 10, 25,
50, 100, 250, 500, 750 Myr, 1, 1.5, 3, 6, 13 Gyr) and metallicities
(0.4 $Z_{\sun}$, 1 $Z_{\sun}$, 2.5 $Z_{\sun}$). To construct the
best fitting model, we perform a non-negative least squares fit with
dust attenuation modeled as an additional free parameter. 

\begin{figure}
\epsscale{0.95}
\plotone{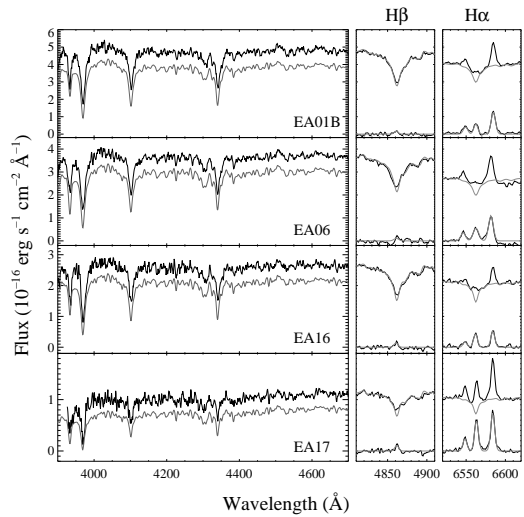}
\caption{Examples of our continuum and line fits.  
  The leftmost panel shows the data (dark) and the best fit continuum 
  model (light, offset for clarity).  The two panels on the right zoom 
  in on the H$\beta$ and H$\alpha$ lines. The upper lines show the data 
  (dark) and the continuum model (light).  The lower lines show the
  continuum subtracted spectrum (dark) and our fit to the nebular
  lines (light).  All three panels have the same vertical scaling.
\label{linefit}}
\end{figure}

After subtracting the best-fitting stellar population model of the
continuum, we fit the nebular emission lines.  Because we are
interested in recovering very weak nebular features, we adopt a
special strategy: we fit all the emission lines with Gaussians
simultaneously, requiring that the Balmer lines (H$\gamma$,
H$\beta$, H$\alpha$) have the same line width and velocity offset,
and likewise for the forbidden lines
([\ion{O}{3}]~$\lambda\lambda4959, 5007$, [\ion{N}{2}]
$\lambda\lambda6548, 6584$, [\ion{S}{2}]~$\lambda\lambda6717, 6731$).
This procedure minimizes the number of free parameters and allows the
stronger lines to constrain the weaker ones.  The
measured H$\alpha$ emission line equivalent widths range 
from 1 to 5 \AA\ for H$\alpha$, and from 0.2 to 1.2~\AA\ for H$\beta$. 
Measurements of lines this weak are only possible because of the high S/N of 
the spectra (10 - 40 per pixel) and our continuum subtraction
techniques\footnotemark.  
We show examples of our continuum and line fits in Figure ~\ref{linefit}.

\footnotetext{The \citet{Tremonti_et_al_2006}
  code allows the line amplitudes to be negative or positive, hence
  there is no overall bias towards detecting lines in emission.}

\begin{figure}
\epsscale{1.0}
\plotone{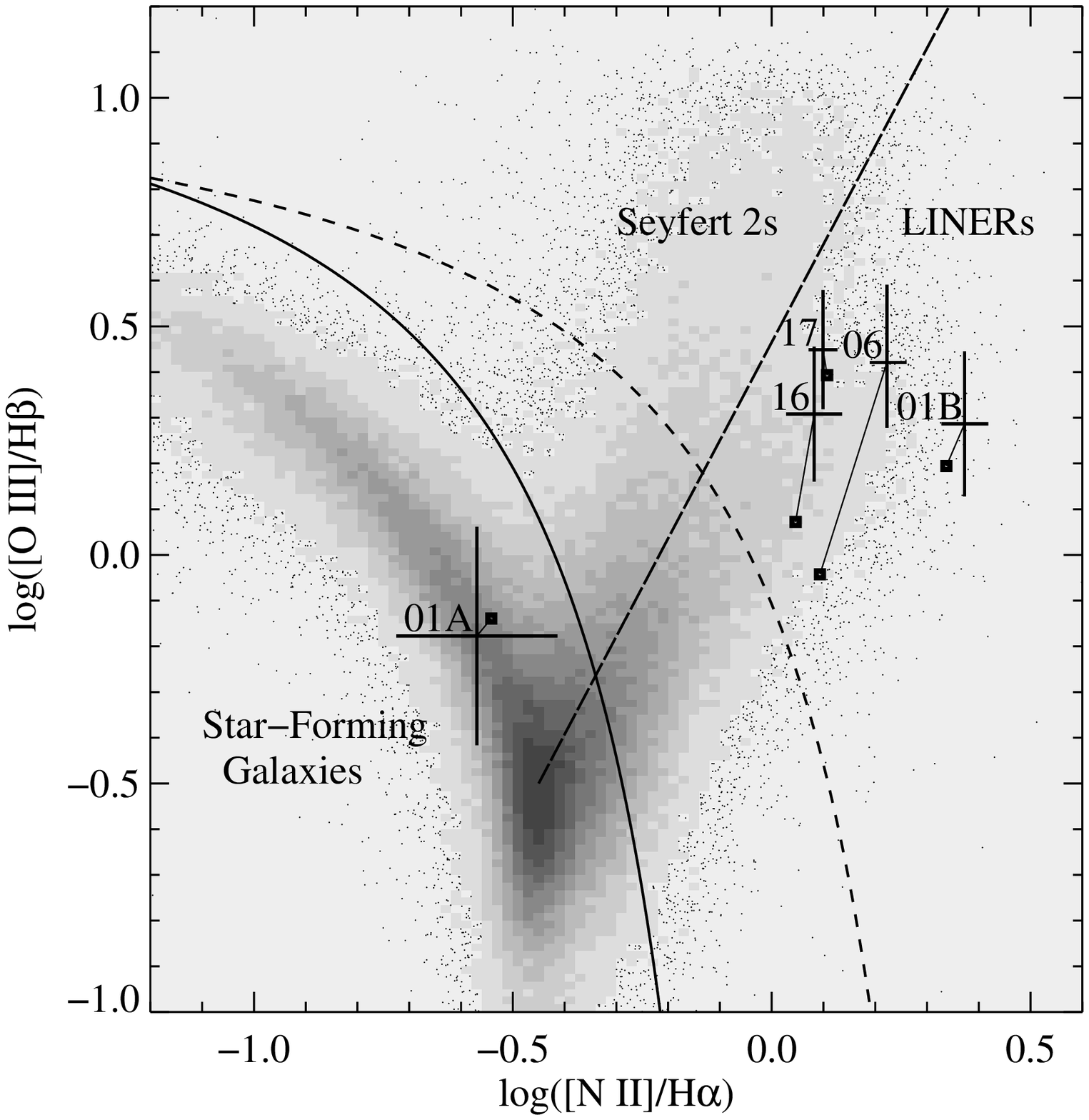}
\caption{AGN diagnostic diagram. The greyscale shows the locus of SDSS
galaxies.  The dashed curve is the AGN division
theoretically calibrated by \citet{Kewley_et_al_2001}; the solid curve
is the empirical division suggested by \citet{Kauffmann_et_al_2003}.
The diagonal dashed line denotes $\Phi = 25\degr$, the
\citeauthor{Kauffmann_et_al_2003} division between Seyfert 2s and
LINERs.  The points with error bars indicate our E+A sample, fit with the
\citet{Bruzual_and_Charlot_2003} models.  
The square points show the effect of fitting the continuum
with the models of \citet{Gonzalez_Delgado_2005} and are an
indication of the systematic error in the measurements.  
\label{bpt}}
\end{figure}

To obtain error estimates for the line fluxes that include errors
in the continuum subtraction, we adopt a bootstrap approach. For each
spectrum, we take the difference between the continuum and the data as
a measure of the error.  We randomly re-sample the errors in bins of
500~\AA, add them to the model of the continuum and nebular lines, and
fit the resultant spectrum with our code.  We do this 1000 times per
galaxy, and use the spread in the measured emission lines as our estimate
of the error.  The technique does not address 
possible systematic errors in the stellar population
models.  Therefore, we perform an identical fitting procedure,
substituting the synthetic spectra of \citet{Gonzalez_Delgado_2005}
for the empirical ones of \citet{Bruzual_and_Charlot_2003}. The
resulting differences in line fluxes are small in EA01A, EA01B, and
EA17, but exceed the random errors in EA06 and EA16 (see Fig.~\ref{bpt}).

The weak nebular emission lines extracted from our spectra provide a
key AGN diagnostic heretofore unavailable for most E+As.  Following
\citet{Baldwin_Phillips_and_Terlevich_1981} and
\citet{Veilleux_and_Osterbrock_1987}, we plot the flux ratios,
[\ion{O}{3}]~$\lambda5007$/H$\beta$ versus \
[\ion{N}{2}]~$\lambda6584$/H$\alpha$, in Figure~\ref{bpt} (widely
referred to as the BPT diagram).  For comparison, we plot 254,548
emission line galaxies from the Sloan Digital Sky Survey (SDSS) Data
Release 4.  The SDSS data form two distinct and remarkably narrow
sequences in this plot, the first corresponding to gas ionized by 
massive, main-sequence stars (the star forming sequence), the second
to gas ionized by other means (the Seyfert/LINER sequence).
\citet{Kewley_et_al_2001} theoretically calibrate a limit
(dashed curve in Fig.~\ref{bpt}) above which galaxies cannot be
explained by any possible combination of parameters in a standard star forming
model.  \citet{Kauffmann_et_al_2003} use SDSS data to empirically
calibrate a limit that more closely adheres to the star forming
galaxy locus (solid curve in Fig.~\ref{bpt}).  

Our E+As occupy an interesting part of the BPT plot.  EA01A falls
squarely on the locus of star forming galaxies, suggesting that there
is minor star formation that was undetected previously due to
dilution by light from larger radii.
In contrast, the other four sources lie well above the star forming locus,
even considering the uncertainties.
\citet{Kauffmann_et_al_2003} quantify the position of AGN in the BPT
plot using a polar scheme centered at the point where the AGN leave
the star forming galaxy locus.  Galaxies are characterized by their
distance $D$ from the origin and by an angle $\Phi$, which is zero
along the positive [\ion{O}{3}]/H$\beta$ axis and increases clockwise.
Our four blue-core E+As have $D=1-1.1$ and $\Phi = 30\degr$-$46\degr$.
\citet{Kauffmann_et_al_2003} classify such galaxies as LINERs.
%

The nature of LINERs \citep{Heckman_1980} remains a puzzle. LINERs may be
low-luminosity AGN (LLAGN) arising from low-rate or low radiative
efficiency accretion onto super-massive black holes \citep{Halpern83,Ferland83}, 
or they may be the product of other mechanisms 
\citep[see][and references therein]{Filippenko_2003} such as photoionization 
by young clusters during the Wolf-Rayet phase \citep{Barth_and_Shields_2000}.
For example, a 4 Myr old star cluster was identified as the dominant ionizing 
source of the LINER nucleus in NGC~4303 \citep{Colina_et_al_2002}, 
but subsequent Chandra imaging revealed a hard X-ray point source indicative 
of an AGN \citep{Jimenez-Bailon_et_al_2003}. 
In the nuclei of our E+As, the Balmer absorption lines constrain the mass 
fraction of $<$ 10 Myr old stars to be less than 0.1\%, ruling out a major 
contribution from young stars. 

A more plausible alternative to ionization by an AGN is ionization by planetary
nebula nuclei that appear in the late-phase evolution of intermediate mass stars.
(Indeed, some LINER galaxies have significant intermediate-age stellar populations 
\citep{Cid_Fernandez_et_al_2004,Gonzalez_Delgado_et_al_2004}.)
However, PNe nuclei are inefficient producers of H$\alpha$ photons
\citep{Taniguchi_et_al_2000}, and the intermediate age stellar mass
required to generate the H$\alpha$ luminosities of our nuclear spectra 
(a few $\times 10^{39}$~erg~s$^{-1}$) is large ($\sim10^{9}$~M$_{\sun}$ 
in the slit aperture).  While such large masses are not ruled out, a LLAGN
provides a more natural explanation. Indeed, many LINERs have nuclear activity
difficult to explain via stellar processes: compact flat-spectrum radio
cores or jets, X-ray point sources with power-law spectra, and broad
Balmer emission \citep[][and references therein]{Ho_2004}.
\citet{Maoz_et_al_2005} found that 80\% of LINERs with compact UV cores
show significant variability, strongly linking LINERs to AGNs.   Thus, the
LINER spectra in our E+As are \emph{likely} signposts of weak AGN
activity, but we acknowledge that intermediate mass stars may provide an
alternate or additional source of gas ionization.

\section{Discussion} 
Using spectral diagnostics, we discover LINERs in the centers of four
E+A galaxies with blue cores.  These LINERs have low luminosities 
similar to those in many nearby early-type galaxies \citep{Ho97}.
The existence of LINERs in E+As, which are typically post-merger,
post-starburst, and bulge-dominated galaxies, is more evidence that
E+As will evolve into early-type galaxies.

If these LINERs are low-luminosity AGN, {\it i.e.}, powered
by black holes,
a more interesting aspect of our discovery is the apparently
simultaneous end of star formation and significant BH growth. 
If we estimate the AGN bolometric luminosity from the [\ion{O}{3}] 
luminosity \citep{Heckman04}, and the BH mass and associated 
Eddington luminosity from the $M_\bullet-\sigma_B$ relation 
\citep{Ferrarese00,Gebhardt00} and Norton et al.'s (2001) velocity dispersion,
the black holes in these E+A's are currently radiating 
at only 0.1 - 0.5 \% of their theoretical maximum luminosity. 
Thus any period of strong AGN activity and significant BH growth,
as might occur during the gas-rich merger \citep{Sanders88}, 
has now ended.  The same is true for the star formation:
after experiencing a burst within the last Gyr, 
these galaxies are now forming stars at rates of 
less than $0.01-0.08\; \rm M_{\sun}\ yr^{-1}$ \citep[derived from the
measured H$\alpha$ luminosities assuming $\rm SFR\; (M_{\sun} \ yr ^{-1})$ = 
$L_{\rm H\alpha} / $ $\rm 1.27\times10^{41}ergs\ s^{-1}$;][]{Kennicutt98}.

These results have implications for the intersection of the small-scale
physics associated with BH evolution
and the large-scale physics that determines the properties of the host
galaxy's bulge.  The coherence of the $M_\bullet-\sigma_B$ relation
\citep{Ferrarese00,Gebhardt00} requires that these small- and
large-scale processes are connected, either causally or through a
third process, particularly in systems where $M_\bullet$ or
$\sigma_B$ is changing by large factors in a short time.  If these LINERs
are LLAGNs that represent the end of BH growth, and the E+A phase marks the
point at which the galactic bulge, enhanced via the galaxy-galaxy
merger and resulting starburst, stops growing, the study of E+As will
provide unique insights into the origin of this relation.

If the cores of these E+As do harbor LLAGN, are the AGN really fading?
\citet{Kauffmann_et_al_2003} find that powerful AGNs (Seyferts with
large $D$ and $\Phi < 25\degr$) lie in massive early-type
galaxies with young stellar populations, whereas weak AGNs (pure
LINERs with large $D$ and $\Phi > 25\degr$) inhabit normal early-type
galaxies.  
Given that the blue core E+As are evolving from late- to early-type
galaxies through the post-starburst phase, it is quite possible that
their AGNs are also in transition (or fading) from the strong
(Seyfert) to the weak (LINER) AGNs.  Finding Seyfert nuclei
among the youngest E+As 
would confirm this conjecture, but our sample is presently too small 
for this type of search.

Is the apparent relationship between the small-scale (LLAGN-sized) and
large-scale (bulge-sized) physics in E+A galaxies causal?  Recent
numerical simulations including BHs suggest that collisions between
galaxies trigger an inflow of gas that causes a strong circumnuclear
starburst \emph{and} fuels BH accretion.  A powerful quasar outflow
removes the gas from the inner region of the merger remnant, quenching
the star formation on a relatively short timescale \cite[$\sim$1
Gyr;][]{Springel05a,Springel05b}.  These models ultimately produce
elliptical galaxies within a few Gyr and, unlike past simulations
without AGN-feedback \citep{Miho_Hernquist94,Miho_Hernquist96},
effectively predict a true E+A phase in which the merger remnant has
no lingering star formation.  It is possible, though not proved, that
the LINERs in our blue core E+As are the relic (fading) AGNs that once
caused the truncation of star formation in these post-starburst
galaxies.  If so, their presence suggests a path by which early-type
galaxies arrive on the $M_\bullet-\sigma_B$ relation and also solves
the long-standing puzzle of what mechanism during the merger ends the
star formation in E+As.

What is the occurrence of LINER spectra in the general post-starburst
galaxy population?  
In recent work, \citet{Yan05} investigate the
emission line properties of a large sample of galaxies drawn from the
SDSS.  They find that AGN are characterized by large [OII]/H$\alpha$
EW ratios and conclude that post-starburst samples defined with an
[OII] EW cut (such as ours) will be biased against AGN.  However, our
high S/N spectroscopy reveals that even E+As with negligible emission
(1-5 \AA\ at H$\alpha$) have line ratios characteristic of AGN.
\citet{Yan05} are able to use the BPT diagram (i.e., our Figure 3) to
classify only 40\% of their post-starburst sample owing to the
weakness of the emission lines.  Of the galaxies they can classify,
$\sim90\%$ harbor AGN.  They suggest that many of the weaker-lined
objects may have LINER-like spectra, which is consistent with our
results.

A final puzzle is the nature of the blue cores, which first led us to
investigate these galaxies.  While our ground-based spectral resolution 
is not ideal for addressing this question, the low emission line luminosity 
of our LINERs and the lack of broad emission lines suggest that the blue
light is not from an AGN continuum.  
It may instead arise from a (fading) circumnuclear starburst, which grew 
in concert with any nuclear activity. Blue cores are common in early-type galaxies 
at higher redshift ($z\gtrsim0.5$), when field spheroids were assembling.
For example, 30\% of the morphologically-selected elliptical galaxies 
in the Hubble Deep Field North have color inhomogeneities, mostly due to 
blue cores \citep{Menanteau2001}. \citet{Treu2005} find that $\sim$ 15\% 
(2/14) of blue-core spheroids in the Great Observatories Origins Deep Survey
North (GOODS-N) field have X-ray detections ($L_X > 10^{42}$ ergs s$^{-1}$), 
suggesting the presence of AGN. Our blue-core, LINER E+As may be local 
examples of a phenomenon common at high redshift.

\section{Conclusions}

We identify four E+A galaxies with blue cores, which are revealed by
our follow-up spectroscopy to have LINER spectra.  The existence of
LINERs, similar to those in elliptical galaxies, is 
more evidence that E+A galaxies, with their
post-starburst spectra, post-merger, gas-poor, bulge-dominated
morphologies, and pressure-supported kinematics
\citep[e.g.,][]{Norton01,Chang01,Yang04}, evolve into normal
early-types.  More interestingly, if LINERs are low-luminosity AGN,
their presence in E+As
suggests that any rapid growth phase of the central AGN has ended in
rough concert with the star formation and therefore that the evolution
of the black hole is tied to that of the galactic bulge.  What is not
clear from our work is whether the coupling between AGN and bulge
evolution is causal, as is suggested by some theoretical models
incorporating AGN feedback \citep{Springel05a,Springel05b}. The study
of a large sample of E+As, including an investigation of any
correlation between AGN strength and the time elapsed since the
starburst, could provide a test of the AGN-feedback hypothesis, as
those models predict that the black hole accretion rate peaks shortly
after the starburst and declines quickly as the merger remnant ages.

\acknowledgements
We thank Felipe Menanteau and the other, anonymous, referee for their helpful 
suggestions. YY and AIZ acknowledge funding from HST grant GO-09781.03-A and 
NSF grant AST 02-06084. CT acknowledges support from HST grant HF-0119201A.



\begin{thebibliography}{}

\bibitem[Baldwin et al.(1981)]{Baldwin_Phillips_and_Terlevich_1981} Baldwin, J.~A., Phillips, M.~M., \& Terlevich, R.\ 1981, \pasp, 93, 5 
\bibitem[Barth \& Shields(2000)]{Barth_and_Shields_2000} Barth, A.~J., \& Shields, J.~C.\ 2000, \pasp, 112, 753 
\bibitem[Bruzual \& Charlot(2003)]{Bruzual_and_Charlot_2003} Bruzual, G.~\& Charlot, S., \mnras, 344, 1000
\bibitem[Chang et al.(2001)]{Chang01} Chang, T., van Gorkom, J.~H., Zabludoff, A.~I., Zaritsky, D., \& Mihos, J.~C.\ 2001, \aj, 121, 1965
\bibitem[Cid Fernandes et al.(2004)]{Cid_Fernandez_et_al_2004} Cid Fernandes, R., et al.\ 2004, \apj, 605, 105 
\bibitem[Colina et al.(2002)]{Colina_et_al_2002} Colina, L., Gonzalez Delgado, R., Mas-Hesse, J.~M., \& Leitherer, C.\ 2002, \apj, 579, 545 
\bibitem[Ferland \& Netzer(1983)]{Ferland83} Ferland, G.~J., \& Netzer, H.\ 1983, \apj, 264, 105 
\bibitem[Ferrarese \& Merritt(2000)]{Ferrarese00} Ferrarese, L., \& Merritt, D.\ 2000, \apjl, 539, L9 
\bibitem[Filippenko(2003)]{Filippenko_2003} Filippenko, A.~V.\ 2003, in Active Galactic Nuclei: from Central Engine to Host Galaxy, ed. S.~Collin, F.~Combes, \& I.~Shlosman (San Francisco: ASP), 387  
\bibitem[Gebhardt et al.(2000)]{Gebhardt00} Gebhardt, K., et al.\ 2000, \apjl, 539, L13 
\bibitem[Gonz{\'a}lez-Delgado et al.(2005)]{Gonzalez_Delgado_2005} Gonzalez-Delgado, R.~M., Cervino, M., Martins, L.~P., Leitherer, C., \& Hauschildt, P.~H.\ 2005, \mnras, 357, 945
\bibitem[Gonz{\'a}lez Delgado et al.(2004)]{Gonzalez_Delgado_et_al_2004} Gonz{\'a}lez Delgado, R.~M.~et al.\ 2004, \apj, 605, 127
\bibitem[Halpern \& Steiner(1983)]{Halpern83} Halpern, J.~P., \& Steiner, J.~E.\ 1983, \apjl, 269, L37 
\bibitem[Heckman(1980)]{Heckman_1980} Heckman, T.~M., 1980, \aap, 87, 152
\bibitem[Heckman et al.(2004)]{Heckman04} Heckman, T.~M., Kauffmann, G., Brinchmann, J., Charlot, S., Tremonti, C., \& White, S.~D.~M.\ 2004, \apj, 613, 109 
\bibitem[Ho et al.(1997)]{Ho97} Ho, L.~C., Filippenko, A.~V., \& Sargent, W.~L.~W.\ 1997, \apj, 487, 568 
\bibitem[Ho(2004)]{Ho_2004} Ho, L.~C., 2004, in Carnegie Observatories Astrophysics Series, Vol 1: Coevolution of Black Holes and Galaxies, ed. L.~C. Ho, (Cambridge: Cambridge Univ.~Press)
\bibitem[Jim{\'e}nez-Bail{\'o}n et al.(2003)]{Jimenez-Bailon_et_al_2003} Jim{\'e}nez-Bail{\'o}n, E.~et al.\ 2003, \apj, 593, 127
\bibitem[Kauffmann et al.(2003)]{Kauffmann_et_al_2003} Kauffmann, G.~et al.\ 2003, \mnras, 346, 1055 
\bibitem[Kennicutt(1998)]{Kennicutt98} Kennicutt, R.~C.\ 1998, \araa, 36, 189 
\bibitem[Kewley et al.(2001)]{Kewley_et_al_2001} Kewley, L.~J., Dopita, M.~A., Sutherland, R.~S., Heisler, C.~A., \& Trevena, J.\ 2001, \apj, 556, 121 
\bibitem[Maoz et al.(2005)]{Maoz_et_al_2005} Maoz, D., Nagar, N.~M., Falcke, H., \& Wilson, A.~W.\ 2005, \apj, 625, 699
\bibitem[Menanteau et al.(2001)]{Menanteau2001} Menanteau, F., Abraham, R. G., \& Ellis, R. S.\ 2001, \mnras, 322, 1
\bibitem[Mihos \& Hernquist(1996)]{Miho_Hernquist96} Mihos, J.~C., \& Hernquist, L.\ 1996, \apj, 464, 641 
\bibitem[Mihos \& Hernquist(1994)]{Miho_Hernquist94} Mihos, J.~C., \& Hernquist, L.\ 1994, \apjl, 425, L13 
\bibitem[Norton et al.(2001)]{Norton01} Norton, S.~A., Gebhardt, K., Zabludoff, A.~I., \& Zaritsky, D.\ 2001, \apj, 557, 150
\bibitem[O'Connell et al.(2005)]{O'Connell05} O'Connell, R.~W.~et al.\ 2005, \apj, 635, 305
\bibitem[Peng et al.(2002)]{Peng02} Peng, C.~Y., Ho, L.~C., Impey, C.~D., \& Rix, H.\ 2002, \aj, 124, 266
\bibitem[Sanders et al.(1988)]{Sanders88} Sanders, D.~B., Soifer, B.~T., Elias, J.~H., Madore, B.~F., Matthews, K., Neugebauer, G., \& Scoville, N.~Z.\ 1988, \apj, 325, 74 
\bibitem[Springel et al.(2005b)]{Springel05b} Springel, V., Di Matteo, T., \& Hernquist, L.\ 2005, \apjl, 620, L79 
\bibitem[Springel et al.(2005a)]{Springel05a} Springel, V., Di Matteo, T., \& Hernquist, L.\ 2005, \mnras, 361, 776 
\bibitem[Taniguchi et al.(2000)]{Taniguchi_et_al_2000} Taniguchi, Y., Shioya, Y., \& Murayama, T.\ 2000, \aj, 120, 1265 
\bibitem[Tremonti et al.(2006)]{Tremonti_et_al_2006} Tremonti, C.~A.,   Brinchmann, J.,  Seibert, M., Bradley, L.\ 2006, in prep.
\bibitem[Treu et al.(2005)]{Treu2005} Treu, T., et al.\ 2005, \apj, 633, 174
\bibitem[Veilleux \& Osterbrock(1987)]{Veilleux_and_Osterbrock_1987} Veilleux, S., \& Osterbrock, D.~E.\ 1987, \apjs, 63, 295 
\bibitem[Yan et al.(2005)]{Yan05} Yan, R., Newman, J.~A., Faber, S.~M., Konidaris, N., Koo, D., \& Davis, M.\ 2005, \apj\ in press.
\bibitem[Yang et al.(2004)]{Yang04} Yang, Y., Zabludoff, A.~I., Zaritsky, D., Lauer, T.~R., \& Mihos, J.~C.\ 2004, \apj, 607, 258 
\bibitem[Yang et al.(2006)]{Yang06} Yang, Y., Zabludoff, A.~I., Zaritsky, D. \& Mihos, J.~C.\ 2006,  in prep.
\bibitem[Zabludoff et al.(1996)]{Zabludoff96} Zabludoff, A.~I.~et al.\ 1996, \apj, 466, 104


\end{thebibliography}
\end{document}